  \documentclass[final,5p,times,twocolumn]{elsarticle}

\usepackage{amssymb,amsmath}
\usepackage{verbatim}

\journal{Physics Letters B}

\begin{document}

\begin{frontmatter}



\title{Admissible higher-spin algebras in flat space}


     \author[first,second]{Dmitry Ponomarev}
\affiliation[first]{organization={I.E. Tamm Theory Department, Lebedev Physical Institute},
            addressline={Leninskiy prospekt 53}, 
            city={Moscow},
            postcode={119991}, 
            country={Russia}}

\affiliation[second]{organization={Institute for Theoretical and Mathematical Physics, Lomonosov Moscow State University},
            addressline={Leninskie Gory, GSP-1}, 
            city={Moscow},
            postcode={119991}, 
            country={Russia}}

\begin{abstract}
We study admissible higher-spin algebras in four-dimensional Minkowski space, namely algebras that admit the tower of massless higher-spin fields as a representation.
Under several simplifying assumptions, we show that these can be of two types, which we refer to as collinear and half-collinear. 
The chiral higher-spin algebra provides an example of the latter.  
For the collinear ansatz considered here, direct solution of the Jacobi identity yields two families of Lie algebras. We also find an associative collinear algebra, which is commutative unless internal degrees of freedom are added.
\end{abstract}




\end{frontmatter}




\section{\label{intro}Introduction}

Difficulties with interactions of massless higher-spin  fields in flat space have long been known \cite{Weinberg:1964ew,Coleman:1967ad}.
Since then many further no-go results within different frameworks have been obtained, see \cite{Bekaert:2010hw} for review. At the same time, there is growing evidence that interacting massless higher-spin theories in flat space do exist, however, the associated scattering amplitudes should involve distributions. 

Indeed, the existing no-go theorems 
do not rule out higher-spin theories with distributional amplitudes. In particular, the Coleman-Mandula theorem  \cite{Coleman:1967ad} explicitly assumes that amplitudes are analytic functions, which are non-vanishing everywhere, except, possibly, some isolated kinematics.
Other no-go results, such as \cite{Weinberg:1964ew,Metsaev:1991mt,Benincasa:2007xk,McGady:2013sga,Ponomarev:2017nrr,Serrani:2026dbs}, either rely on the requirement of consistent factorisation -- that is the requirement that tree-level amplitudes have only first order poles and the associated residues factorise into lower-point subamplitudes -- or on the equivalent requirements. 
For distributional amplitudes this analysis needs to be revisited. Moreover, one can anticipate additional subtleties related to  summations over spins.
Finally, some no-go results, such as \cite{Joung:2013nma,Roiban:2017iqg}, rely on the covariant descriptions of massless higher-spin theories in terms of Fronsdal fields \cite{Fronsdal:1978rb}, which do not capture even all possible cubic interactions, see e. g. \cite{Bengtsson:2014qza,Conde:2016izb,Ponomarev:2022vjb}.

At the same time, in the last few years, several examples of distributional higher-spin amplitudes were found.
 In particular, a distributional four-point amplitude in conformal higher-spin theory \cite{Tseytlin:2002gz,Segal:2002gd} was found in \cite{Joung:2015eny}.
 Similarly, four-point amplitudes for higher-spin theories in AdS computed holographically are distributional in Mellin space \cite{Taronna:2016ats,Bekaert:2016ezc}.
A similar result also holds for chiral higher-spin theories \cite{Metsaev:1991mt,Ponomarev:2016lrm}.
  Namely, by requiring invariance with respect to the global  chiral higher-spin algebra candidate chiral higher-spin amplitudes were constructed in \cite{Ponomarev:2022atv}.
These also feature delta functions in addition to those required by momentum conservation. Yet another example is provided by continuous-spin fields for which distributions emerge already at the level of cubic interactions 
\cite{Metsaev:2017cuz,Basile:2026ntj}.

It is important to note that distributional amplitudes also appear for lower-spin theories. Self-dual Yang-Mills theory and self-dual gravity are integrable, so by the Coleman-Mandula theorem the associated amplitudes cannot be analytic and non-trivial. As was found recently \cite{Guevara:2026qzd,Guevara:2026qwa} these are, indeed, non-trivial and supported on specific half-collinear kinematics. Remarkably, as argued there, despite the unusual analytic structure, these amplitudes can be computed from the associated local actions. 
This result  suggests that chiral higher-spin amplitudes from \cite{Ponomarev:2022atv} can  be also  found by the direct application of the Feynman rules or other equivalent techniques. 
This issue was recently discussed in \cite{Serrani:2026azw}.

In this letter our goal is to explore the possibilities for massless higher-spin theories in flat space beyond the chiral case \cite{Metsaev:1991mt,Ponomarev:2016lrm}.
More specifically, we will focus on the analysis of suitable global higher-spin algebras. By exploring  higher-spin theories at the level of symmetries we avoid the necessity of dealing with the potential non-locality issues, which are not yet clear how to resolve. At the same time, higher-spin symmetries  is a very powerful tool as these typically fix the theory almost completely. For example, higher-spin amplitudes in AdS space \cite{Colombo:2012jx,Didenko:2012tv,Gelfond:2013xt} and chiral higher-spin amplitudes in flat space \cite{Ponomarev:2022atv} are both fixed up to an overall factor for each number of external lines by global symmetries alone. Besides that, higher-spin symmetries can be used to define the holographically dual theories by following a straightforward procedure of \cite{Eastwood:2002su}. This idea was used to construct a holographic dual of flat-space chiral higher-spin theories in \cite{Ponomarev:2022ryp,Ponomarev:2022qkx}.

 Higher-spin algebras in flat space were studied previously, see e. g.  \cite{Bekaert:2006us,Bekaert:2008sa,Joung:2013nma,Sleight:2016xqq,Campoleoni:2021blr}. These works, however, either  rely on the Fronsdal fields or produce algebras, which do not satisfy one or another crucial consistency condition. Among these is the requirement that a higher-spin algebra should admit massless higher-spin fields as a representation. This requirement turns out to be very constraining, as was pointed out  in \cite{Konstein:1989ij} when studying possible extensions of the higher-spin algebra in AdS space \cite{Fradkin:1986ka}. We will use the terminology of \cite{Konstein:1989ij} and refer to higher-spin algebras that admit massless higher-spin representations as admissible.

Below we will explore higher-spin algebras in four-dimensional flat space focusing on the requirement that these admit massless higher-spin fields as a representation. 
The concrete implementation of this requirement used by us is discussed  in section  \ref{sec:2}. 
We then solve the resulting constraints in section \ref{sec:3}. First, we find that admissible higher-spin algebras can be of one of the two types, which we refer to as collinear and half-collinear. Then, focusing on the collinear case, we find two families of higher-spin Lie algebras by directly solving the Jacobi identity. 
In section \ref{sec:4} we discuss our results. In particular, we identify the Lie  algebras found and discuss their associative counterparts.
Finally, we conclude in section
\ref{sec:conclusions}.

\section{Setup}
\label{sec:2}

In the present paper we will deal with massless fields in four-dimensional Minkowski space. In this context it is convenient to use $sl(2,\mathbb{C})$ spinors. For  conventions 
on $sl(2,\mathbb{C})$  spinors and for various useful formulas we refer the reader to  \cite{Elvang:2013cua}.

Consistency conditions we need to solve are as follows. Firstly, the global higher-spin algebra Lie bracket should be antisymmetric $[\varepsilon_1,\varepsilon_2]= -[\varepsilon_2,\varepsilon_1]$ and satisfy the Jacobi identity
\begin{equation}
\label{17feb1}
[\varepsilon_1,[\varepsilon_2,\varepsilon_3]] +
[\varepsilon_2,[\varepsilon_3,\varepsilon_1]]+
[\varepsilon_3,[\varepsilon_1,\varepsilon_2]]=0.
\end{equation}
Next, massless fields $\Phi$ should transform $\delta_{\varepsilon}\Phi = T(\varepsilon)\Phi$ in some representation of this algebra, that is
\begin{equation}
\label{17feb2}
[T(\varepsilon_1),T(\varepsilon_2)]=T([\varepsilon_1,\varepsilon_2]).
\end{equation}
Higher-spin algebras satisfying (\ref{17feb2}) will be  referred to as admissible.

To make the problem more manageable, below we will make a few simplifying assumptions. The first  assumption concerns the spectrum of the theory. 
In principle, there exist higher-spin theories with different spectra. For example, three-dimensional higher-spin theories in AdS space may have  finite spectra, see e. g. 
\cite{Campoleoni:2010zq}. There are also various options for the spectra in chiral higher-spin theories \cite{Ponomarev:2016lrm,Ponomarev:2017nrr,Ponomarev:2024jyg,Serrani:2025owx}. Below we will assume that a theory involves the complete tower of bosonic massless higher-spin fields, 
moreover, each field appears only once. This is the case, in particular, for higher-spin theories with established holographic duals \cite{Sezgin:2002rt,Klebanov:2002ja,Ponomarev:2022ryp,Ponomarev:2022qkx}.

It is convenient to encode the associated spectrum of on-shell fields in terms of a single function $\Phi(\lambda,\bar\lambda)$ of $sl(2,\mathbb{C})$ spinors $\lambda$ and $\bar\lambda$. A field of helicity $h\in \frac{1}{2}\mathbb{Z}$ satisfies
\begin{equation}
\label{16feb3}
H\Phi(\lambda,\bar\lambda) = h\;\Phi(\lambda,\bar\lambda), \quad  H\equiv \frac{1}{2}\left({N}_{\bar\lambda}- 
N_\lambda \right),
\end{equation}
where
\begin{equation}
\label{16feb3x1}
{N}_{\bar\lambda} \equiv  \bar\lambda^{\dot\alpha}\frac{\partial}{\partial \bar\lambda^{\dot\alpha}} , \qquad    N_\lambda\equiv 
\lambda^{\alpha}\frac{\partial}{\partial \lambda^{\alpha}}.
\end{equation}
Strictly speaking, the zero-momentum representations are not  massless fields, so $\Phi(\lambda,\bar\lambda)$ are functions on $\mathbb{C}^2/\{0\}$.

 In covariant descriptions based of Fronsdal fields each spin-$s$ gauge field, $s=|h|$, comes together with a global symmetry algebra parameter, which can be identified with a rank-$(s-1)$ traceless Killing tensor, see \cite{Fradkin:1986ka,Bekaert:2006us,Joung:2013nma,Sleight:2016xqq,Campoleoni:2021blr}. Quite conveniently, in four dimensions these can also be combined into a single function $\varepsilon(y,\bar{y})$ of $sl(2,\mathbb{C})$ spinors. In these terms parameters associated with spin $s$ satisfy
 \begin{equation}
\label{16feb3x2}
\frac{1}{2}({N}_{\bar{y}}+ 
N_y )\varepsilon(y,\bar{y}) = (s-1)\;\varepsilon(y,\bar{y}),
\end{equation}
where $N_{\bar{y}}$ and $N_y$ are defined analogously to (\ref{16feb3x1}).
 It needs to be emphasised that fields and parameters belong to different functional classes: while fields are given by square normalisable functions, global symmetry algebra parameters are polynomial in $sl(2,\mathbb{C})$ spinors. 

Next, we should implement the requirement that $\Phi(\lambda,\bar\lambda)$ describes massless fields in four-dimensional Minkowski space. 
We would like to do that in a way that is general enough to cover the chiral higher-spin theory case, which is associated with the following two subtleties.
Firstly, for chiral higher-spin theories the complete global symmetry algebra does not have the Poincare algebra as a subalgebra, instead, it is replaced with a certain version of the Maxwell algebra \cite{Ponomarev:2022atv,Ponomarev:2022ryp}. At the same time, the latter algebra acts on massless higher-spin fields in a representation for which translations still commute, reproducing the standard Poincare algebra action
\begin{equation}
\label{16feb1}
\begin{split}
T(P_{\alpha\dot\alpha}) \Phi(\lambda,\bar\lambda)&=-\lambda_\alpha\bar\lambda_{\dot\alpha} \Phi(\lambda,\bar\lambda),\\
T(\bar J_{\dot\alpha\dot\beta}) \Phi(\lambda,\bar\lambda) &= i\left(\bar\lambda_{\dot \alpha}\frac{\partial}{\partial\bar\lambda^{\dot\beta}}+\bar\lambda_{\dot \beta}\frac{\partial}{\partial\bar\lambda^{\dot\alpha}}\right)\Phi(\lambda,\bar\lambda),\\
T(J_{\alpha\beta}) \Phi(\lambda,\bar\lambda) &=i\left(\lambda_{\alpha}\frac{\partial}{\partial \lambda^\beta}+
\lambda_{\beta}\frac{\partial}{\partial \lambda^\alpha} \right) \Phi(\lambda,\bar\lambda)
\end{split}
\end{equation}
of translations $T(P)$ and of (anti-)self-dual components of the Lorentz algebra $T(J)$ and $T(\bar{J})$.
Accordingly, in what follows, we will only assume that the complete global symmetry algebra of a theory has a set of generators, collectively denoted by $g$, that act on higher-spin fields as  (\ref{16feb1}). Secondly, in the chiral higher-spin case the complete global symmetry algebra of the theory is larger than the chiral higher-spin algebra. Namely, the theory enjoys Lorentz symmetry, but the chiral higher-spin algebra contains only its (anti)-self-dual part. In what follows we will not assume that the higher-spin algebra gives the complete global symmetry algebra of a theory.

The complete global symmetry algebra is also a Lie algebra, while on-shell fields form its representation. In particular, one has
\begin{equation}
\label{17feb3}
\begin{split}
&[\varepsilon_1,[\varepsilon_2,g]] +
[\varepsilon_2,[g,\varepsilon_1]]+
[g,[\varepsilon_1,\varepsilon_2]]=0,\\
&[T(\varepsilon),T(g)]=T([\varepsilon,g]).
\end{split}
\end{equation}
These conditions impose additional constraints on consistent higher-spin algebras.

In order to implement (\ref{17feb3}), it would be beneficial if the adjoint action $[\varepsilon,g]$ of $g$ on  the higher-spin generators $\varepsilon$ was  known. 
As we mentioned above, in Fronsdal's approach global symmetry parameters are given by traceless Killing tensors, which, in particular, entails 
\begin{equation}
\label{28jul1}
\begin{split}
[P_{\alpha\dot\alpha},\varepsilon] = y_{\alpha}\frac{\partial}{\partial \bar{y}^{\dot\alpha}} \varepsilon, \qquad ( N_{\bar y} - N_y) \varepsilon >0,\\
[P_{\alpha\dot\alpha},\varepsilon] = \bar{y}_{\dot\alpha}\frac{\partial}{\partial y^{\alpha}} \varepsilon, \qquad ( N_{\bar y} - N_y) \varepsilon <0,\\
[P_{\alpha\dot\alpha},\varepsilon] = 0, \qquad ( N_{\bar y} - N_y) \varepsilon =0.
\end{split}
\end{equation}
In contrast to that, for chiral higher-spin theories one has
\begin{equation}
\label{28jul2}
\begin{split}
[P_{\alpha\dot\alpha},\varepsilon] = y_{\alpha}\frac{\partial}{\partial \bar{y}^{\dot\alpha}} \varepsilon
\end{split}
\end{equation}
irrespective of the homogeneity degrees of $\varepsilon$ in spinors.
Transformations (\ref{28jul1}), (\ref{28jul2}) do not match, and,
 in agreement with that, chiral higher-spin theories 
 are not captured by the standard local formulation in terms of Fronsdal fields \cite{Bengtsson:2014qza,Conde:2016izb,Ponomarev:2022vjb}.
Therefore, the adjoint action of the (deformed) Poincare algebra on the global higher-spin algebra parameters appears to depend on the specific off-shell formalism used and may change from theory to theory.

In order to proceed, we will make the second assumption. Namely, we will assume that there is a map $O$ that intertwines the action of the complete global symmetry algebra on the on-shell fields and on the algebra parameters in the sense that
\begin{equation}
\label{28jul3}
[g,O\cdot \Phi] = O\cdot T(g)\Phi, \qquad [\varepsilon,O\cdot \Phi] = O\cdot T(\varepsilon)\Phi.
\end{equation}
Such a map exists both for higher-spin theories in AdS space and for chiral higher-spin theories with Metsaev's coupling constants \cite{Metsaev:1991mt}.
In both cases it can be given by the Fourier transform in the half of the spinor variables \cite{Didenko:2009td,Didenko:2012tv,Ponomarev:2022atv}
\begin{equation}
\label{30jul1}
(O\cdot\Phi )(y,\bar{y})\propto \int d^2\bar{\lambda} e^{i[\bar{y}\bar\lambda]}\Phi(\lambda,\bar\lambda)\big|_{\lambda = y}.
\end{equation}

A few comments are in order. First of all, we would like to emphasise that the existence of such an intertwining map does not mean that the associated representations are equivalent, since global algebra parameters and on-shell fields belong to different functional classes. Indeed, after the Fourier transform the transformed fields remain square normalisable, while $\varepsilon$ are polynomials. Thus, even the mathematically accurate formulation of the assumption (\ref{28jul3}) is a subtle issue. 
At the same time, it is very useful practically. 
Namely, once the intertwining assumption (\ref{28jul3}) is adopted, the problem of admissible higher-spin algebras simplifies in the following way. Equations (\ref{17feb1}) and (\ref{17feb2}) are now equivalent. The same applies to two equations in (\ref{17feb3}). Therefore, in what follows we will only solve the Jacobi identity (\ref{17feb1}) and the first equation in (\ref{17feb3}). In the latter equation for $[\varepsilon,g]$ we will use (\ref{16feb1}), that is we will assume that $\varepsilon$ is taken in the basis in which the (deformed) Poincare algebra acts the same way as it does on on-shell fields.
In this context, the functional class issue does not seem to be relevant. Indeed, in (\ref{17feb1}), (\ref{17feb2}) and (\ref{17feb3}) both $\Phi$ and $\varepsilon$ appear 
as functions the structure constants are integrated with. Irrespective of the particular functional class,
equality for a sufficiently rich class of test functions implies equality of the corresponding distributional kernels.

Before closing this section, we will make a few remarks on how (\ref{28jul3}) can be potentially justified. As we mentioned, this property holds for higher-spin theories in AdS and for chiral higher-spin theories with Metsaev's coupling constants. Both these theories have holographic duals, so it is suggestive to try to justify the intertwining property by holography. For holographic higher-spin theories the spectrum of the higher-spin algebra is given by $\varepsilon \in {\rm End}(V)\sim V \otimes V^*$, where $V$ is the representation of the holographically dual theory, while $V^*$ is its dual. At the same time, the spectrum of the on-shell fields is given by the Flato-Fronsdal theorem \cite{Flato:1978qz}, $\Phi \in V\otimes V$. Unitarity of the dual theory allows one to establish the equivalence $V \sim V^\dagger = \bar{V}^*$, so ${\rm End}(V) \sim V\otimes \bar{V}$. This line of thought, though, seems encouraging, still, does not allow us to map ${\rm End}(V) $ to $ V\otimes V$, as required. We would also like to note that (\ref{28jul3}), apparently, holds for individual massless fields, thus, the underlying holographic picture is not required. In particular, massless fields and the associated global symmetry parameters satisfy the same Casimir equations. This means that both representations can be realised by the same set of differential operators acting on functions that belong to different functional classes. Related discussions can be found in e.g. \cite{Iazeolla:2008ix,Bekaert:2017khg,Basile:2018dzi}. 
At present, we do not have a general and rigorous justification of  property (\ref{28jul3}), so we will keep it as an assumption. We will be content with the fact that the
assumption (\ref{28jul3}), though, may, in principle, rule out some additional possibilities,  still allows us to construct admissible higher-spin algebras, see below. 

\section{Solving the consistency conditions}
\label{sec:3}

In this section we will solve for admissible higher-spin algebras within the setup detailed above. We will first solve the Poincare invariance condition given by the first equation in (\ref{17feb3}). We will then focus on a particular class of solutions found and then proceed with the Jacobi identity (\ref{17feb1}).

\subsection{Collinear and half-collinear algebras}

Let the commutator be given by a general integral kernel $L$
\begin{equation}
\label{30jul2}
\begin{split}
[\varepsilon_1,\varepsilon_2](\lambda_3,\bar\lambda_{3})= \int d^2\lambda_1d^2\bar\lambda_1d^2\lambda_2 d^2\bar\lambda_2 L(\lambda_i,\bar\lambda_i)\\
 \varepsilon_1(\lambda_1,\bar{\lambda}_1)
\varepsilon_2(\lambda_2,\bar{\lambda}_2),
\end{split}
\end{equation}
where $i=1,2,3$.
For the Poincare algebra generators acting as in (\ref{16feb1}), the first equation (\ref{17feb3}) has the form of the Poincare invariance condition imposed on the three-point amplitude  $L$.

This problem is standard, see e.g. \cite{Benincasa:2007xk}, and the solution typically proceeds as follows.
Momentum conservation implies 
\begin{equation}
\label{30jul3}
[ij]\langle j k\rangle =0, 
\end{equation}
where $i$, $j$ and $k$ are the particle numbers from the set $\{1, 2, 3\}$.
In order to satisfy (\ref{30jul3}) one has to choose out of the following options. The first option is 
to choose solutions, which are supported on the branch for which the undotted spinors are collinear, while the dotted ones are not
\begin{equation}
\label{30jul4}
\langle ij \rangle =0, \qquad [ij]\ne 0.
\end{equation}
The second option is the opposite one: the dotted spinors are collinear, while the undotted ones are not
\begin{equation}
\label{30jul5}
\langle ij \rangle \ne 0, \qquad [ij]= 0.
\end{equation}
Amplitudes with support  (\ref{30jul4}) and (\ref{30jul5}) require  dotted and undotted spinors to be independent, which can be achieved by going to  the split signature or by making momenta complex.
Finally, there is a more exotic third option, for which spinors of both types are collinear simultaneously
\begin{equation}
\label{30jul6}
\langle ij \rangle=  0, \qquad [ij]= 0.
\end{equation}
This equation can be satisfied with real momenta in the Lorentzian signature, which is what we will assume in what follows when dealing with this case.
We will refer to higher-spin algebras with the structure constants supported on (\ref{30jul4}) or (\ref{30jul5}) as half-collinear, while the last case (\ref{30jul6}) will be referred to as collinear.

An example of a half-collinear higher-spin algebra is provided by the chiral higher-spin algebra \cite{Ponomarev:2017nrr,Skvortsov:2022syz,Sharapov:2022faa,Ponomarev:2022atv}. In a suitable basis the associated structure constants read
\begin{equation}
\label{30jul7}
L = k \sinh ({l[12]}) \delta^2(\bar\lambda_1+\bar\lambda_2-\bar\lambda_3) \delta^2(\lambda_2-\lambda_3)\delta^2(\lambda_1-\lambda_3),
\end{equation}
where $k$ and $l$ are free parameters. Similarly, one can consider a half-collinear higher-spin algebra with the structure constants given by the complex conjugate of (\ref{30jul7}).
We briefly note that a naive parity-invariant completion of (\ref{30jul7}), which consists of combining it with its complex conjugate does not work, as it results in the violation of the Jacobi identity.

In what follows we will focus on collinear higher-spin algebras. Collinearity can be implemented via a combination of delta functions
\begin{equation}
\label{26aug1}
\delta^2(\lambda_1-\alpha_1\lambda_3)\delta^2(\lambda_2-\alpha_2\lambda_3)
\delta^2(\bar\lambda_1-\bar\alpha_1\bar\lambda_3)\delta^2(\bar\lambda_2-\bar\alpha_2\bar\lambda_3),
\end{equation}
which is manifestly Lorentz invariant if $\alpha$'s do not transform. Implementing momentum conservation with another delta function, we end up with the following ansatz for the Lie bracket
\begin{equation}
\label{28jul1x1}
\begin{split}
&[\varepsilon_1,\varepsilon_2](\lambda_3,\bar\lambda_{3})= \int d\alpha_1d\alpha_2 d\bar{\alpha}_1d\bar{\alpha}_2 f(\alpha_1,\bar{\alpha}_1,\alpha_2,\bar\alpha_2)\\
& \qquad \delta(1-\alpha_1\bar{\alpha}_1-\alpha_2\bar{\alpha}_2) \varepsilon_1(\alpha_1\lambda_3,\bar{\alpha}_1\bar{\lambda}_3)
\varepsilon_2(\alpha_2\lambda_3,\bar{\alpha}_2\bar{\lambda}_3).
\end{split}
\end{equation}

It needs to be remarked that (\ref{28jul1x1}) does not provide the most general solution of the Poincare invariance condition  with the collinear support (\ref{30jul6}).
Namely, equation (\ref{28jul1x1}) covers the sector in which the two fields are evaluated directly on the fully collinear locus. More general Poincare-invariant kernels with the same support may contain derivatives of the delta functions in the directions transverse to this locus. We do not consider these terms here.

\subsection{Collinear solutions to the Jacobi identity}

It is convenient to pass to new variables 
\begin{equation}
\label{11aug2}
\begin{split}
\lambda_\alpha = u^{-\frac{1}{2}}(2\omega)^{\frac{1}{2}}  (-z,1)_\alpha, \qquad \bar\lambda_{\dot \alpha}= u^{\frac{1}{2}}(2\omega)^{\frac{1}{2}} (-\bar{z},1),
\end{split}
\end{equation}
where $\omega$ is the so-called energy scale, while $z$ and $\bar{z}$ are the standard coordinates on the celestial sphere, see e.g. \cite{Raclariu:2021zjz} for review. In (\ref{11aug2}) we also introduced a pure phase variable
$u\bar{u}=1$, which allows us to combine all higher-spin contributions into a single generating function 
\begin{equation}
\label{26au2}
\varepsilon(\omega,z,\bar{z},u) =\sum_{h\in\mathbb{Z}} \varepsilon^h(\omega,z,\bar{z}) u^h,
\end{equation}
where $h$ is helicity, which we assume to be integer. In these terms, after straightforward manipulations the bilinear product (\ref{28jul1x1}) becomes
\begin{equation}
\label{11aug17}
\begin{split}
&[\varepsilon_1,\varepsilon_2]^{h_3}(\omega,z,\bar{z})=
 \sum_{h_1+h_2=h_3} \int_0^1 d\pi a_{h_1,h_2}(\pi)\\
 &\qquad\qquad\qquad \qquad
\varepsilon_1^{h_1}(\omega\pi,z,\bar{z})\varepsilon_2^{h_2}(\omega(1-\pi),z, \bar{z}),
\end{split}
\end{equation}
where up to an overall normalization
\begin{equation}
\label{11aug16}
\begin{split}
&a_{h_1,h_2}(\pi)
=
- \int_0^1  d\varphi_1  d\varphi_2 e^{-2i h_1\varphi_1}
 e^{-2ih_2\varphi_2} \\
& \qquad f(\sqrt{\pi} e^{i\varphi_1},\sqrt{\pi} e^{-i\varphi_1},\sqrt{1-\pi} e^{i\varphi_2},\sqrt{1-\pi} e^{-i\varphi_2}).
 \end{split}
\end{equation}
The Jacobi identity then leads to
\begin{equation}
\label{29jul1}
\begin{split}
 &\frac{1}{\pi_1+\pi_2}a_{h_3,h_1+h_2}(1-\pi_1-\pi_2)a_{h_1,h_2}\left(\frac{\pi_1}{\pi_1+\pi_2}\right)\\
&+\frac{1}{1-\pi_1}
a_{h_1,h_2+h_3}(\pi_1)a_{h_2,h_3}\left(\frac{\pi_2}{1-\pi_1}\right)\\
&+\frac{1}{1-\pi_2}
a_{h_2,h_3+h_1}(\pi_2)a_{h_3,h_1}\left(\frac{1-\pi_1-\pi_2}{1-\pi_2}\right)
=0,
\end{split}
\end{equation}
and the skew symmetry of the Lie bracket translates as
\begin{equation}
\label{15feb1}
a_{h_1,h_2}(\pi)= - a_{h_2,h_1}(1-\pi).
\end{equation}

The Jacobi identity (\ref{29jul1}) gives a system of algebraic equations, which need to be solved for all integer $h_1$, $h_2$ and $h_3$.
In order to solve it we note that, once the skew symmetry (\ref{15feb1}) is  taken into account, as $\pi_2\to 0$, all arguments in (\ref{29jul1}) go to $\pi_1$ or $0$. Thus, by differentiating (\ref{29jul1}) with respect to $\pi_2$ and then sending $\pi_2$ to zero, one can find differential equations for $a(\pi_1)$ featuring derivatives of $a$ at zero as parameters.
We solved these equations assuming that $a(\pi)$ can be expanded in a Laurent series at $\pi=0$ and
its principal part contains finitely many terms.

The solution along these lines is rather technical and it will be presented elsewhere. Here we just quote the end result for the general solution within the Laurent series ansatz and the generic non-vanishing assumption specified below. 
The first branch of solutions is given by
\begin{equation}
\label{31aug31}
\begin{split}
a_{h_1,h_2}(\pi)&=\frac{q_{h_1}q_{h_2}}{q_{h_1+h_2}}(1-\pi)^{\kappa h_2 -2}\pi^{\kappa h_1-2}\\
&\big((\gamma h_2+\beta)\pi-(\gamma h_1+\beta)(1-\pi)\big),
\end{split}
\end{equation}
where $q_h\ne 0$, $\kappa$, $\beta$ and $\gamma$ are arbitrary parameters.
The second branch is 
\begin{equation}
\label{31aug31x1}
\begin{split}
a_{h_1,h_2}(\pi)=\frac{q_{h_1}q_{h_2}}{q_{h_1+h_2}}(1-\pi)^{\kappa h_2 -1}\pi^{\kappa h_1-1}
(h_2-h_1).
\end{split}
\end{equation}

In the process of solving the Jacobi identity we  assumed that $a_{h_1,h_2}(\pi)$ are nonzero for generic helicity pairs, otherwise, other solutions exist. For example, in (\ref{31aug31}) one can keep only $a_{0,0}(\pi)$ non-vanishing. Similarly, one can consistently set to zero all $a_{h_1,h_2}(\pi)$ for which both $h_1$ and $h_2$ are non-vanishing.
The systematic analysis of such truncations requires a laborious case by case study, see \cite{Serrani:2025owx} for an analogous analysis of consistent sectors of chiral higher-spin theories. We leave it for future research.

\section{Discussion}
\label{sec:4}

Let us briefly discuss the algebras found and their properties. For brevity, we will ignore the spectator $z$ and $\bar{z}$ variables.
First, we define
\begin{equation}
\label{26aug5}
 F^h(r)  \equiv \int_0^\infty e^{-r\omega} \omega^{\kappa h-2}q_h\varepsilon^h(\omega).
\end{equation}
With  $F^h(r)$ combined into a generating function
\begin{equation}
\label{26aug6}
F(r,v) \equiv \sum_{h\in \mathbb{Z}} F^h(r) v^{\frac{\gamma}{\beta} h +1}
\end{equation}
the structure constants (\ref{31aug31}) lead to  the bracket
\begin{equation}
\label{26aug7}
\beta^{-1}[F_1,F_2] = {\partial_v F_1} {\partial_r F_2} -  {\partial_v F_2} {\partial_r F_1}.
\end{equation}
This is just the usual Poisson bracket for functions $F(r,v)$, which feature $v$ to powers $\frac{\gamma}{\beta} \mathbb{Z} +1$. For the special case $\gamma=0$ 
introducing 
\begin{equation}
\label{26aug6x1}
\hat{F}(r,t) \equiv \sum_{h\in \mathbb{Z}} \hat{F}^h(r) t^h
\end{equation}
one finds
\begin{equation}
\label{26aug6x2}
\beta^{-1}[\hat{F}_1,\hat{F}_2] = { \hat{F}_1} {\partial_r \hat{F}_2} -  { \hat{F}_2} {\partial_r \hat{F}_1},
\end{equation}
which is a loop extension of the algebra of vector fields in $r$.
 In turn, for the special case $\beta=0$ it is more natural to combine $F^h(r)$ into
\begin{equation}
\label{26aug8}
\tilde{F}(r,\varphi) \equiv \sum_{h\in \mathbb{Z}} F^h(r) e^{ih\varphi}.
\end{equation}
Then, the structure constants (\ref{31aug31}) entail the Poisson algebra of functions on the cylinder
\begin{equation}
\label{26aug9}
i  \gamma^{-1} [\tilde{F}_1,\tilde{F}_2] = {\partial_{\varphi} \tilde{F}_1} {\partial_r \tilde{F}_2} -  {\partial_{\varphi} \tilde{F}_2} {\partial_r \tilde{F}_1}.
\end{equation}
In a similar manner one can show that (\ref{31aug31x1}) provides a realisation of a current algebra  of the Witt algebra.

The existence of the holographic dual requires that higher-spin algebras are associative rather than Lie algebras \cite{Eastwood:2002su}. It is, thus, interesting to explore whether the Lie algebras found can be promoted to associative algebras. The Poisson bracket cannot be obtained as a commutator algebra of an associative algebra. At the same time, it can be deformed to the Weyl-Moyal bracket, which can then be promoted to an associative algebra. It turns out that for all cases listed above such a deformation violates Poincare invariance of the bracket. For example, the Moyal-Weyl deformation of (\ref{31aug31}) with $\beta=0$ results in the associative algebra with the product
\begin{equation}
\label{27aug1}
\begin{split}
b^{(\hbar)}_{h_1,h_2}(\omega,\pi)=\,&\frac{\gamma}{\hbar}\frac{q_{h_1}q_{h_2}}{q_{h_1+h_2}}(1-\pi)^{\kappa h_2 -2}\pi^{\kappa h_1-2}\\
&\frac{1}{\omega}
\exp \left(\frac{\hbar \omega}{2}\left[h_2\pi - h_1(1-\pi)\right] \right),
\end{split}
\end{equation}
where $\hbar $ is the deformation parameter. The presence of the energy scale $\omega$ in (\ref{27aug1}) is inconsistent with Poincare invariance, see (\ref{11aug17}).

To confirm these results we revisited our analysis focusing on the associativity instead of the Jacobi identity. By proceeding along the same lines as before we found the following structure constants
\begin{equation}
\label{27aug2}
\begin{split}
b_{h_1,h_2}(\pi)=\frac{q_{h_1}q_{h_2}}{q_{h_1+h_2}}(1-\pi)^{\kappa h_2 -1}\pi^{\kappa h_1-1}.
\end{split}
\end{equation}
The associated product is commutative, which agrees with the previous discussion.
An obvious way to avoid commutativity is to make algebra generators  matrix-valued.

Finally, we briefly mention how the BMS algebra \cite{Bondi:1962px,Sachs:1962wk} emerges in our construction. For simplicity, we will focus on the $\beta=0$ algebra with the commutator in the canonical basis given by
(\ref{26aug9}). For elements of the form 
\begin{equation}
\label{27aug4}
A = -\frac{\varphi}{\gamma} f(z,\bar z)
\end{equation}
the associated adjoint action gives
\begin{equation}
\label{27aug3} 
\delta_A\tilde{F} \equiv [A,\tilde{F}] = {i}f(z,\bar{z})\partial_r \tilde{F}.
\end{equation}
It needs to be remarked, that $A$ in (\ref{27aug4}) is not globally defined if $\varphi$ is assumed to be constrained to a circle $[0,2\pi)$.
In terms of the original variables $(\omega,z,\bar{z},u)$ transformation (\ref{27aug3}) translates into
\begin{equation}
\label{27aug5} 
\delta_A \varepsilon =-i\omega f(z,\bar{z})\varepsilon 
\end{equation}
and reproduces the action of supertranslations on massless fields. 
In fact, as it is not hard to see, the general collinear product (\ref{28jul1x1}) is BMS invariant without the necessity to introduce soft supermomenta.
The emergence of supertranslations makes collinear higher-spin algebras promising candidate symmetries that may underly free-theory examples of Carrollian or celestial holography, see e.g. \cite{Strominger:2017zoo,Raclariu:2021zjz,Bekaert:2025kjb,Nguyen:2025zhg,Ruzziconi:2026bix} for review.

\section{Conclusions}
\label{sec:conclusions}

Our work can be regarded as a step towards the construction of an extension of the Coleman-Mandula theorem which would not rely on analyticity
and non-vanishing of the S-matrix for almost all kinematics. 
The necessity of such an extension is motivated by higher-spin theories for which amplitudes are expected to contain distributions. With the analyticity and non-triviality assumptions relaxed, we are left with very basic quantum field theory symmetry requirements, such as the requirement that global symmetries are given by Lie algebras and on-shell fields transform in their representations. Despite the fact that these requirements may appear weak, for  algebras as rich as higher-spin algebras   these turn out to be very constraining. 

Having made a few simplifying assumptions, we found that in a proper basis higher-spin algebra structure constants satisfy the same constraints as Poincare-invariant three-point amplitudes for massless fields. As a result, flat space higher-spin algebras split into two classes, which we referred to as collinear and half-collinear. Half-collinear higher-spin structure constants have the kinematics of massless amplitudes, which are typically considered in the literature. To the best of our knowledge,  amplitudes with fully collinear kinematics were not discussed before.
An example of a half-collinear higher-spin algebra is provided by the chiral higher-spin algebra. In the present paper we focused on the analysis of the collinear  case. 

When studying collinear higher-spin algebras, we considered an ansatz (\ref{28jul1x1}) for the structure constants, which does not cover all Poincare-invariant billinear maps supported on the collinear kinematics. Even within this limited framework we found  two families of higher-spin Lie algebras.
By suitable changes of variables these algebras can be connected to various versions of the Poisson-bracket algebra,
to a loop extension of an algebra of vector fields in one dimension and to a current algebra of the Witt algebra.
We also carried out the analysis of admissible associative higher-spin algebras along these lines. We found that under the assumptions specified in the text, the only admissible associative algebra is commutative, unless extended with internal degrees of freedom. 

Thus, despite our limited framework, we found some positive results indicating that collinear higher-spin algebras is a promising framework. It would be interesting to revisit our analysis with the most general collinear structure constants taken into account. Besides that, it would be important to refine and justify our intertwining assumption (\ref{28jul3}), relating the on-shell field and adjoint representations of the higher-spin algebra. More generally, it would be important to clarify the mechanism of the emergence of global symmetries for massless theories in a formalism-independent fashion.  

Finally, we would like to mention that, as we found, the collinear higher-spin algebra setting naturally leads to the BMS symmetry. This suggests that, in contrast to the chiral higher-spin algebra, that leads to holography with the underlying Maxwell symmetry \cite{Ponomarev:2022ryp},
 collinear higher-spin algebras may lead to higher-spin holography of a more standard -- such as celestial or Carrollian -- type, see 
 \cite{Campoleoni:2017mbt,Campoleoni:2020ejn,Campoleoni:2021blr,Himwich:2021dau,Ren:2022sws,Bekaert:2022ipg,Monteiro:2022xwq,Hu:2023geb,Tran:2025xbt,Campoleoni:2025bhn,Serrani:2025oaw,Banerjee:2026olu} for other works exploring higher-spin symmetries and holography in flat space  from complementary directions.

\section*{Acknowledgements}
We are grateful to V.~Didenko, N.~Misuna and E.~Skvortsov for stimulating discussions and interesting comments.
The work on sections 1-3 was supported by RSF grant 26-12-00240. 

\paragraph{Declaration of generative AI and AI-assisted technologies in the manuscript preparation process}
During the preparation of this work, the author used OpenAI ChatGPT (GPT-5.6 Sol) to assist with algebraic consistency checks and language editing. After using this tool, the author reviewed, edited and independently verified the material and takes full responsibility for the content of the article.

\appendix



\biboptions{sort&compress}
\bibliographystyle{elsarticle-num} 
\bibliography{collinear}






\end{document}